\documentclass[a4paper, 11pt]{article}
\usepackage[left=19mm, right=19mm, bottom=35mm, top=35mm]{geometry}
\pdfoutput=1
\pdftrue

\usepackage{amsmath}
\usepackage[pdftex]{graphicx}

\usepackage[T2A]{fontenc}
\usepackage[utf8]{inputenc}
\usepackage[english]{babel}
\usepackage{multicol}
\usepackage{colortbl} 

\usepackage{geometry}
\usepackage{amsfonts}
\usepackage{amsmath}
\usepackage{amssymb}
\usepackage{ntheorem}
\usepackage{latexsym}
\usepackage{units}
\usepackage{textcomp}
\usepackage{booktabs}
\usepackage{hyperref}

\usepackage{xcolor}
\hypersetup{
    colorlinks,
    linkcolor={red!50!black},
    citecolor={blue!50!black},
    urlcolor={blue!80!black}
}

\everymath{\displaystyle}
\usepackage{color}
\everymath{\displaystyle}

\renewcommand{\phi}{\varphi}

\usepackage{url}

\begin{document}


\vspace{-20mm}
\begin{center}

\section*{\Large The upper limit on the \boldmath{$K^+ \to \pi^0\pi^0\pi^0e^+\nu$} ~decay}
\end{center}
\vspace{2mm}
\begin{center}
\begin{minipage}{1.0\linewidth}
  {\center \large \textsc{The OKA collaboration}\\}\vspace{-2mm}
\end{minipage}
\end{center}
\begin{center}
\begin{minipage}{1.0\linewidth}
 \center{
  \textsc
  A.~V.~Kulik{${}^{a}$},
  S.~N.~Filippov,
  E.~N.~Gushchin,
  A.~A.~Khudyakov,\\
  V.~I.~Kravtsov,
  Yu.~G.~Kudenko{${}^{b,c}$},
  A.~Yu.~Polyarush
 }\vspace{-3mm}
 \center{\small 
   \textsc{(Institute for Nuclear Research -- Russian Academy of Sciences, 117312 Moscow, Russia),} 
 }\vspace{-1mm}
 \center{
  \textsc
  A.~V.~Artamonov,
  S.~V.~Donskov,
  A.~P.~Filin,
  A.~M.~Gorin,
  A.~V.~Inyakin,
  G.~V.~Khaustov,\\
  S.~A.~Kholodenko,
  V.~N.~Kolosov,
  A.~K.~Konoplyannikov,
  V.~F.~Kurshetsov,
  V.~A.~Lishin,
  M.~V.~Medynsky,\\
  V.~F.~Obraztsov,
  A.~V.~Okhotnikov,
  V.~A.~Polyakov,
  V.~I.~Romanovsky,
  V.~I.~Rykalin,\\
  A.~S.~Sadovsky,
  V.~D.~Samoylenko,
  I.~S.~Tiurin,
  V.~A.~Uvarov,
  O.~P.~Yushchenko
 }\vspace{-3mm}
 \center{\small 
   \textsc{(NRC "Kurchatov Institute"${}^{}_{}{}^{}$-${}^{}_{}{}^{}$IHEP, 142281 Protvino, Russia),} 
 }\vspace{-1mm}
 \center{
  \textsc
  V.~N.~Bychkov, 
  G.~D.~Kekelidze,
  V.~M.~Lysan,
  B.~Zh.~Zalikhanov
 }\vspace{-3mm}
 \center{\small 
   \textsc{(Joint Institute of Nuclear Research, 141980 Dubna, Russia)}\\
 }\vspace{-1mm}
\end{minipage}
\end{center}

{
\footnotesize
\line(1,0){170}\\
\vspace{-1mm}${}$\hspace{0.8cm}${}^{a}$~e-mail: alexkulik@ihep.ru\\ 
\vspace{-1mm}${}$\hspace{0.8cm}${}^{b}$~Also at National Research Nuclear University (MEPhI), Moscow, Russia\\
\vspace{-1mm}${}$\hspace{0.8cm}${}^{c}$~Also at Institute of Physics and Technology, Moscow, Russia\\
}
\vspace{-4mm}
\begin{center}
\begin{minipage}{0.09\linewidth}
~
\end{minipage} 
\begin{center}
\begin{minipage}{0.83\linewidth}
{ 
  \rmfamily
      {\bf Abstract}
      A search for the $K^{+} \to \pi^{0}\pi^{0}\pi^{0}e^+\nu$ decay is performed by the OKA
  collaboration. The search is based on $3.65 \times 10^9 ~ K^+$ decays. No signal is observed.
  The upper limit set is $BR(K^{+} \to \pi^{0}\pi^{0}\pi^{0}e^+\nu) < 5.4\times 10^{-8} ~ 90\%$ CL,
  65 times lower than the one currently listed by PDG.
}\vspace{3mm}
{\\  {\bf Keywords} {Kaon decays~$\cdot$~experimental results}}
\end{minipage}
\end{center}
\begin{minipage}{0.09\linewidth}
~
\end{minipage}
\end{center}
\vspace{-2mm}



\section{Introduction}\label{SectInitro}

The rare $K^+ \to \pi^0\pi^0\pi^0e^+\nu$ (Ke5) decay has not been observed so far. Due to the limited phase space 
the expected BR(Ke5) calculated within Chiral Perturbation Theory is very low, $10^{-11} - 10^{-12}$
\cite{Theor-ITEP, ChPT-1994, ChPT}. On the other hand, the limited phase space favours $\pi\pi$-scattering in the final state which may increase
the BR considerably. For example, a pion atom, {\it pionium} ($A_{2\pi}$), may form with subsequent decay
  into $\pi^0\pi^0$:
$\pi^+\pi^- \to A_{2\pi} \to \pi^0\pi^0$\cite{pionium-Th,pionium-Ex}. Since the pionium decay length $c\tau \sim 10^{-4}$cm is far beyond
the nuclear interaction scale, a 4-body decay $K^+ \to A_{2\pi} \pi^0 e^+ \nu$ should be considered in this case rather than the 5-body one.
This circumstance would increase the phase space of the decay $\sim 10^6$ times\cite{ChPT-1994,ChPT}
with subsequent decay $A_{2\pi} \to \pi^0\pi^0$ ($\approx 100\%$ probability) resulting in $\pi^0\pi^0\pi^0e^+\nu$
final state.

The goal of this study is to considerably improve
the current upper limit on the decay:

$BR(K^{+} \to \pi^{0}\pi^{0}\pi^{0}e^+\nu) < 3.5 \times 10^{-6}$ \cite{Bolotov}.
The upper limit is set by normalization to
$K^{+} \to \pi^{0} e^+\nu$ (Ke3) decay and the systematics is cross-checked by measuring the BR of
$K^{+} \to \pi^{0}\pi^{0}e^+\nu$  (Ke4) decay.

\section{The data and the analysis procedure}

The {\bf OKA} experiment located at NRC ''Kurchatov Institute''-$^{}$IHEP in Protvino (Russia) is dedicated to the
study of kaon decays using the decay in flight technique. 
A secondary kaon-enriched hadron beam is obtained by an RF separation with the Panofsky scheme.
The beam momentum is 17.7~GeV with the kaon content of about 12.5\% and intensity up to $5\times 10^{5}$ kaons per
U-70 accelerator spill.
The OKA setup comprises two magnetic spectrometers, one for the beam and another one for secondary particles, 
along with an 11~m long decay volume equipped with a guard system.
It is complemented with the GAMS and BGD electromagnetic calorimeters \cite{BGDref1982, GAMSref1985}, the GDA
hadron calorimeter \cite{GDAref1986} and the muon identification system.
The setup also has two Cherenkov counters to separate the beam $K^+$ from $p,\pi^+$ and a wide-aperture
4-channel threshold  Cherenkov counter to separate secondary $e^{\pm}$ from $\mu^{\pm},\pi^{\pm}$.
More details on the OKA setup can be found in \cite{OKA_status_2009,  OKA_KmuHnu_EPJC}.
Three runs of data recorded in 2012, 2013 and 2018 are analysed in search for
$K^{+} \to \pi^{0}\pi^{0}\pi^{0}e^+\nu$ decay.
To calculate the detection efficiency the Monte Carlo (MC) events of Ke3, Ke4 and Ke5 decays are generated
with Geant-3.21 \cite{Geant321} program comprising a realistic description of the setup.
The number of generated MC events is $\sim$ 10 times larger than that of the recorded data sample.
For the estimation of the background to the selected data set, a sample of the Monte Carlo events
is generated with six main decay channels of charged kaon
($\mu^+\nu$, $\pi^{+}\pi^{0}$, $\pi^{0}e^{+}\nu$, $\pi^{0}\mu^{+}\nu$, $\pi^{+}\pi^{0}\pi^{0}$, $\pi^{+}\pi^{+}\pi^{-}$)
mixed accordingly to their branching fractions.
The MC events are passed through full OKA reconstruction procedures.
Every MC event is assigned a weight $w \sim |M|^2$ where M is the matrix element of the decay.

\section{Selection of $Ke3, Ke4, Ke5$ events }

\begin{figure*}[!ht]
\begin{center}
\includegraphics[width=0.3\textwidth]{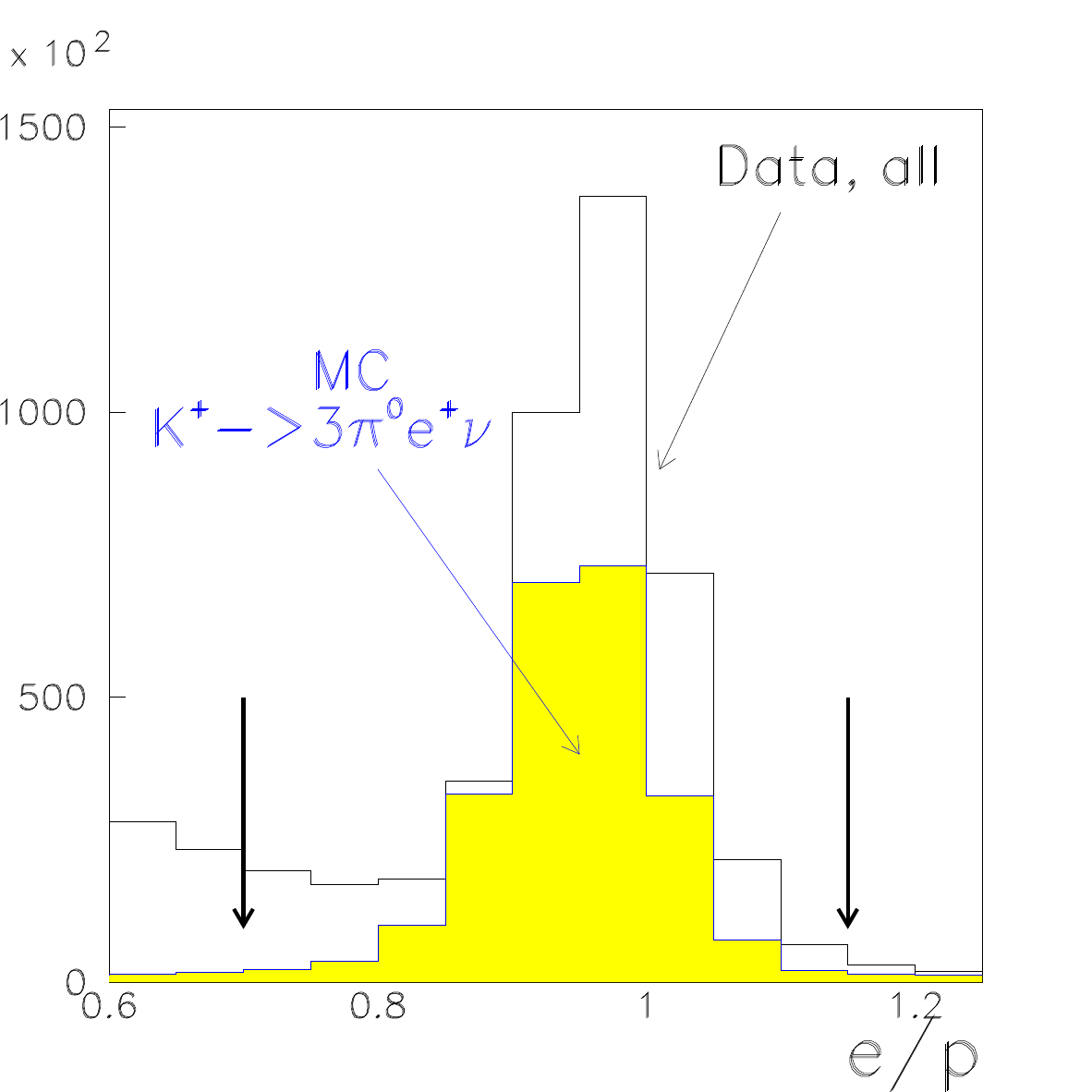}
\includegraphics[width=0.3\textwidth]{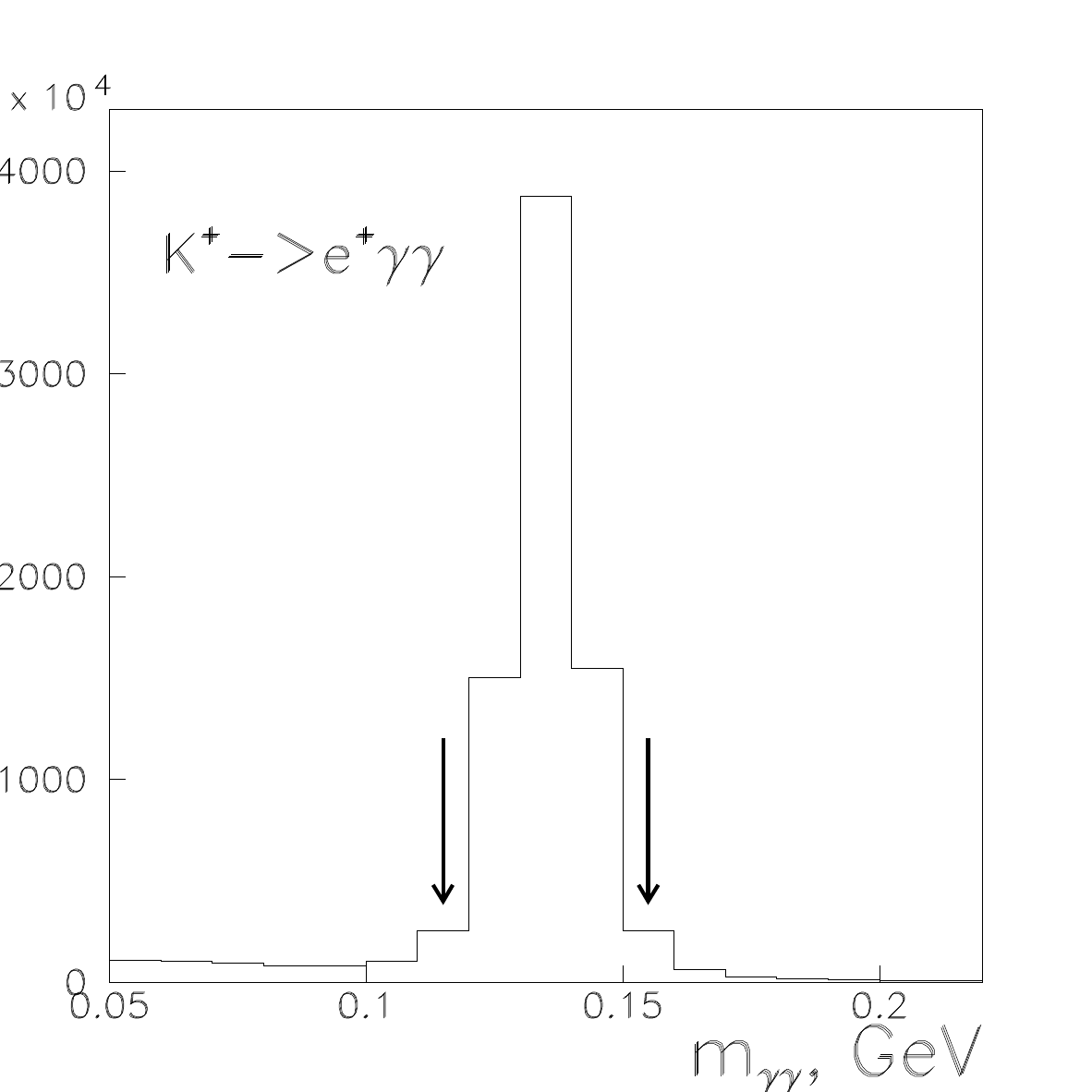}
\includegraphics[width=0.3\textwidth]{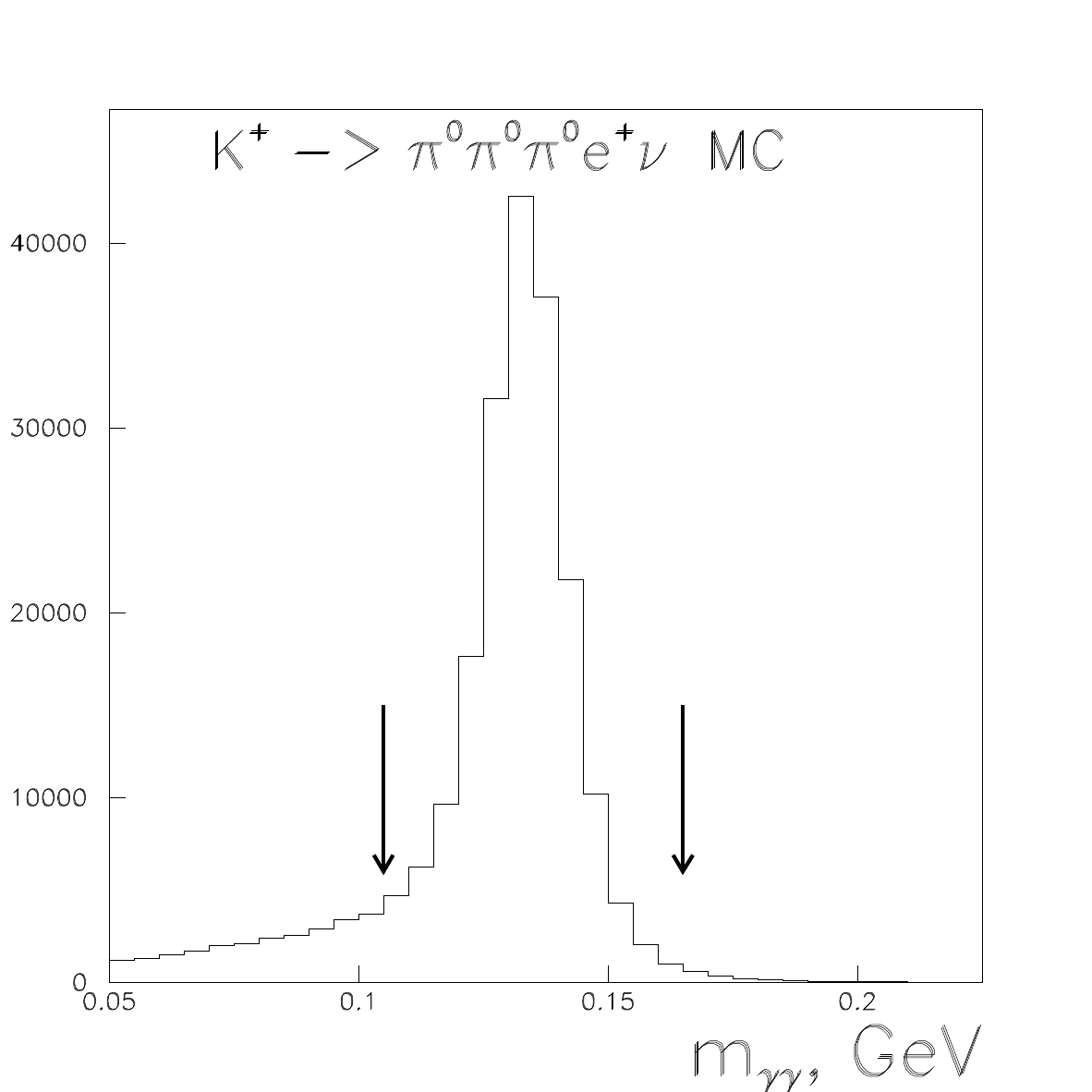}
\caption{\label{cuts} Left: e/p ratio for electron, middle: hard cut for $\pi^0$, right: soft cut for $\pi^0$.}
\end{center}
\end{figure*}

We select  events with a single secondary track identified as $e^+$ and with 1, 2 or 3 of $\pi^0$s found.
The positively charged track is required to have a matching shower in GAMS with e/p ratio within $\pm 3 \sigma$ of its
nominal value (fig.\ref{cuts}(left)). The combinations of $\gamma$s are exhaustively searched through in order to
identify the one with the maximum number of pairs ($\pi^0$s) meeting the condition
  
\begin{equation}
\sum_{i} (m_{\gamma\gamma} - m_{\pi^0})_i^2 < R_{\pi}^2.
\label{pir-def}
\end{equation}

The value of the $R_{\pi}$ parameter will be specified later.
The implementation of this general procedure for the 3 individual decays is somewhat different. 
The ample statistics in $K^+ \to \pi^0e^+\nu$ and
$K^+ \to \pi^0\pi^0e^+\nu$ allows to make the selection criteria more tight and to impose more cuts to further suppress the backgrounds.
On the contrary, there is no signal of $K^+ \to \pi^0\pi^0\pi^0e^+\nu$ decay even with loose cuts thus allowing for increase of
the detection efficiency and rendering the analysis stable against possible MC imperfections. Therefore we support 2 basic criteria sets:
``hard'' ones for $K^+ \to \pi^0\pi^0e^+\nu$ and ``soft'' ones for the searched rare decay
$K^+ \to \pi^0\pi^0\pi^0e^+\nu$.
Such a paradoxial choice is driven by the background-free environment for $K^+ \to \pi^0\pi^0\pi^0e^+\nu$.
We keep track of $K^+ \to \pi^0e^+\nu$ events with soft {\bf and} hard cuts separately for the
normalization of $K^+ \to \pi^0\pi^0\pi^0e^+\nu$  and $K^+ \to \pi^0\pi^0e^+\nu$ respectively.
The differences between hard and soft cuts are as follows.

\begin{itemize}
\item Electron ID in hard cuts is based solely on GAMS: the track projection to the GAMS face is required
  to match one of the GAMS showers as described above. 
Soft cuts use GAMS in the same way plus also allow for tracks with no matching GAMS shower (missing GAMS) but confirmed by 
the wide-aperture Cherenkov counter. The detection efficiency benefits considerably from this feature
because in $K^+ \to \pi^0\pi^0\pi^0e^+\nu$ decay with its soft $e^+$ spectrum the tracks are missing the
GAMS often.
\item Search for $\pi^0$s. 
  The cutoff parameter $R_{\pi}$ in (\ref{pir-def}) is set to $R_{\pi}=0.02$~GeV for hard cuts and $R_{\pi}=0.03$~GeV
  for soft cuts (figs.\ref{cuts}(middle),(right)).
\item Hard cuts require exactly 2 or 4 $\gamma$ with $E_{\gamma}>0.5$~GeV prior to $\pi^0$ search per (\ref{pir-def}).
  Soft cuts search for $\pi^0$s the events with $n_{\gamma} \geq 6$, $E_{\gamma}>0.3$~GeV sometimes finding $3\pi^0$s in a
  $7\gamma$ or even a $8\gamma$ event.
\item Hard cuts require no showers in the GDA hadronic calorimeter and excatly 2 segments of the $e^+$ track:
  one upstream and one downstram the analysing magnet. Soft cuts allow GDA (noise) showers and extra segments of
  the track.
\end{itemize}

\section{Observation of the decays}

Some individual selection criteria are added for each decay at the late stage of analysis.

\subsection{$K^+ \to \pi^0e^+\nu$}

\begin{figure*}[!ht]
\begin{center}
\includegraphics[width=0.3\textwidth]{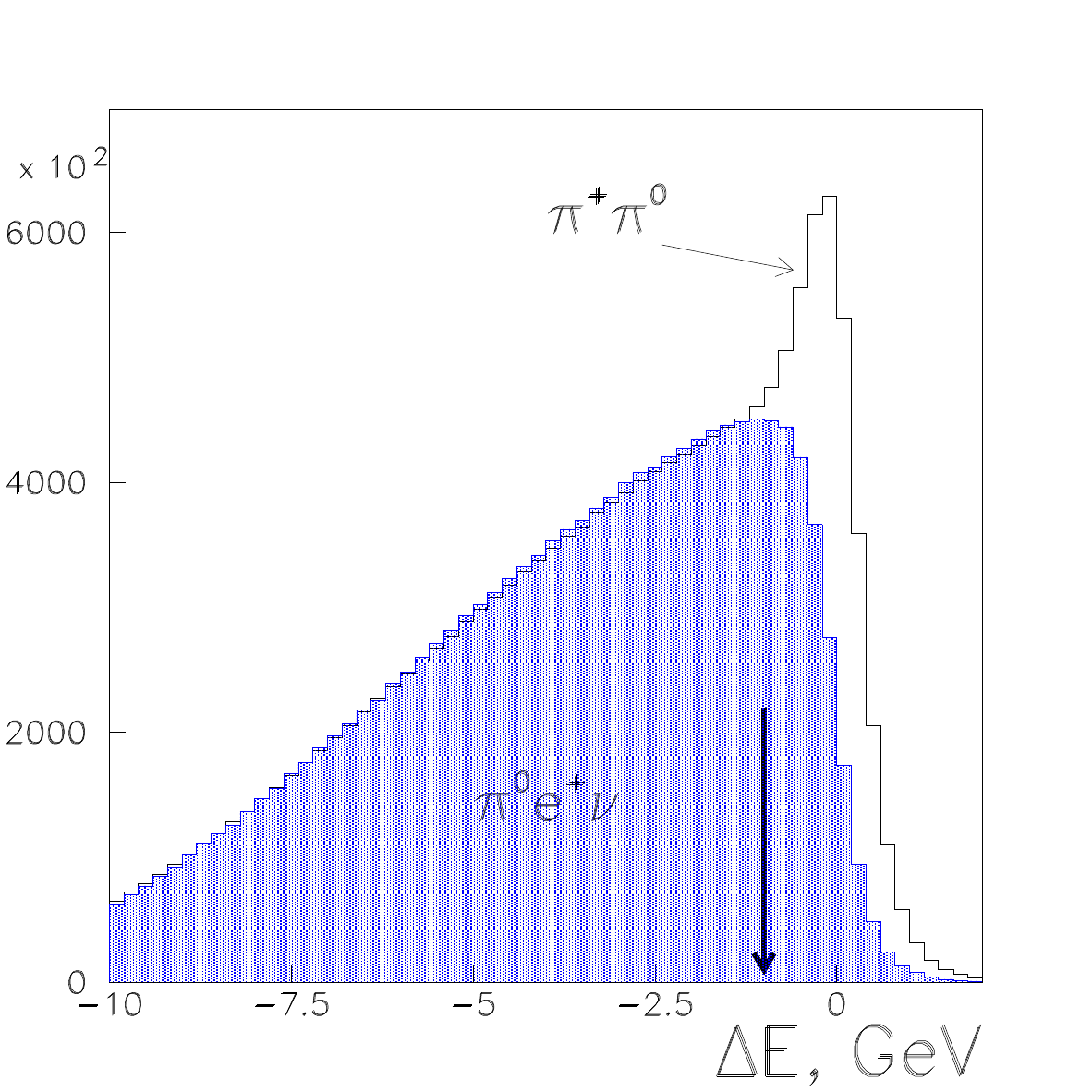}
\includegraphics[width=0.3\textwidth]{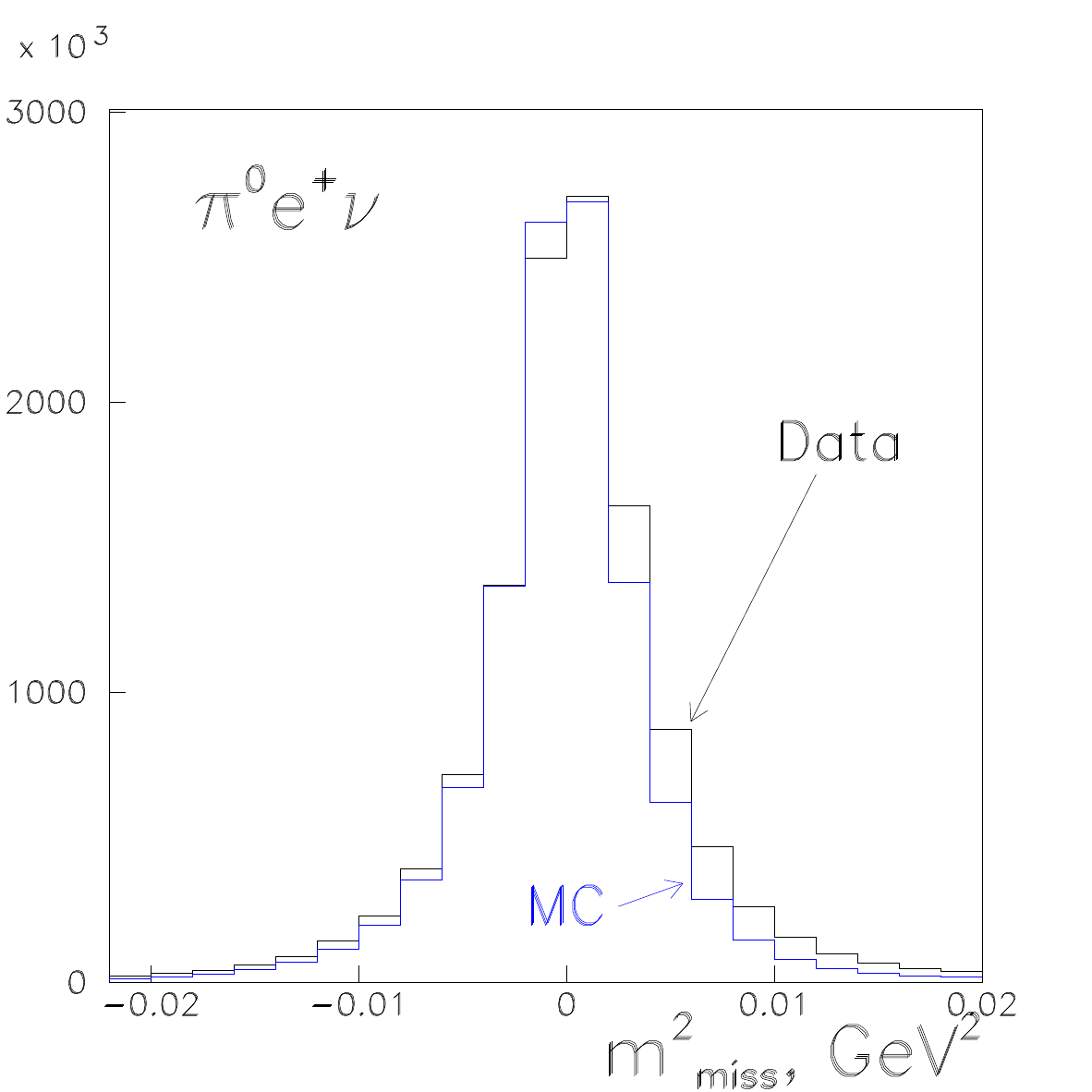}
\caption{\label{ke3} Observation of $K^+ \to \pi^0e^+\nu$ decay.
  Left: energy balance, the cut $\Delta E<-1$~GeV applied to suppress the background
  from $\pi^+\pi^0$; right: missing mass squared, data and MC.}
\end{center}
\end{figure*}

The energy balance in the event is defined as $\Delta E = E_{det} - E_{b}$ where $E_{det}$ is the sum of energies
of the detected particles ($e^+,\pi^0s)$ and $E_{b}$ is the beam energy. Since $\nu$ is not detected,
the decays with $\nu$ should have $\Delta E<0$. With the $\Delta E<-1$~GeV cut 
imposed on top of hard selection criteria to suppress the background from $\pi^+\pi^0$ (fig.\ref{ke3}(left))
a clear peak in the missing mass squared containing $\sim 8.4 \times 10^6$ events is observed (fig.\ref{ke3}(right)).
The MC estimates the background to this sample to $\sim 1\%$.

\subsection{$K^+ \to \pi^0\pi^0e^+\nu$}

\begin{figure*}[!ht]
\begin{center}
\includegraphics[width=0.3\textwidth]{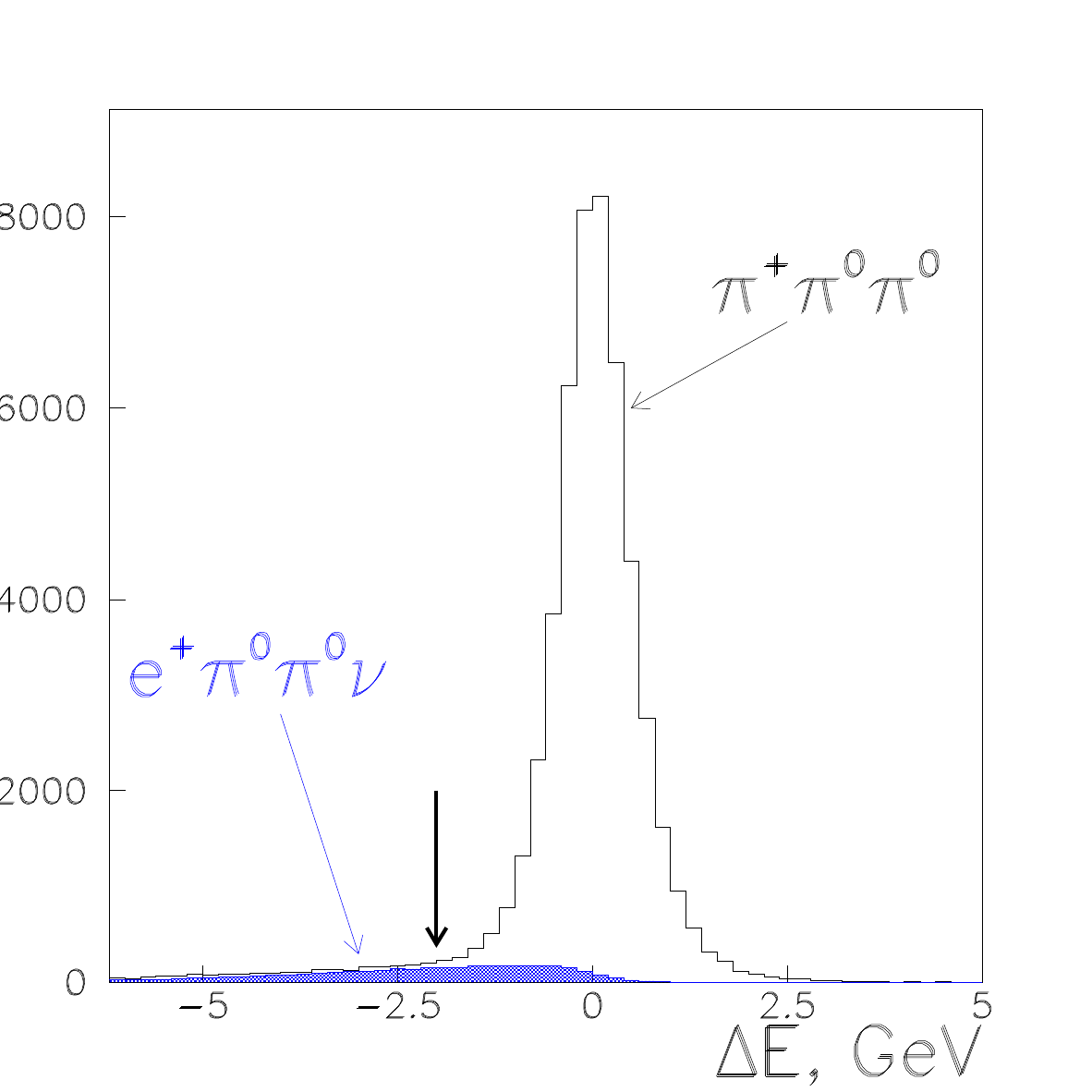}
\includegraphics[width=0.3\textwidth]{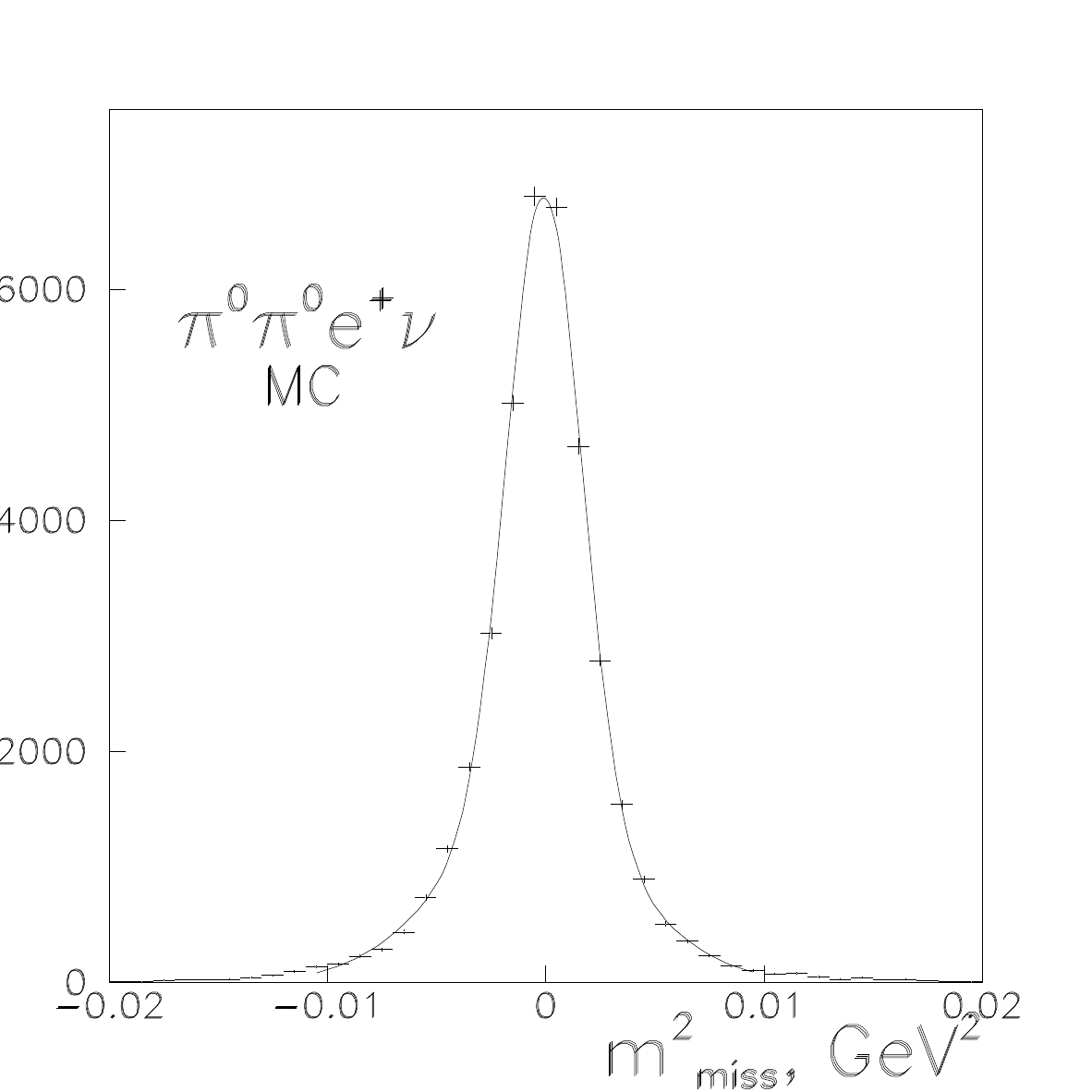}
\includegraphics[width=0.3\textwidth]{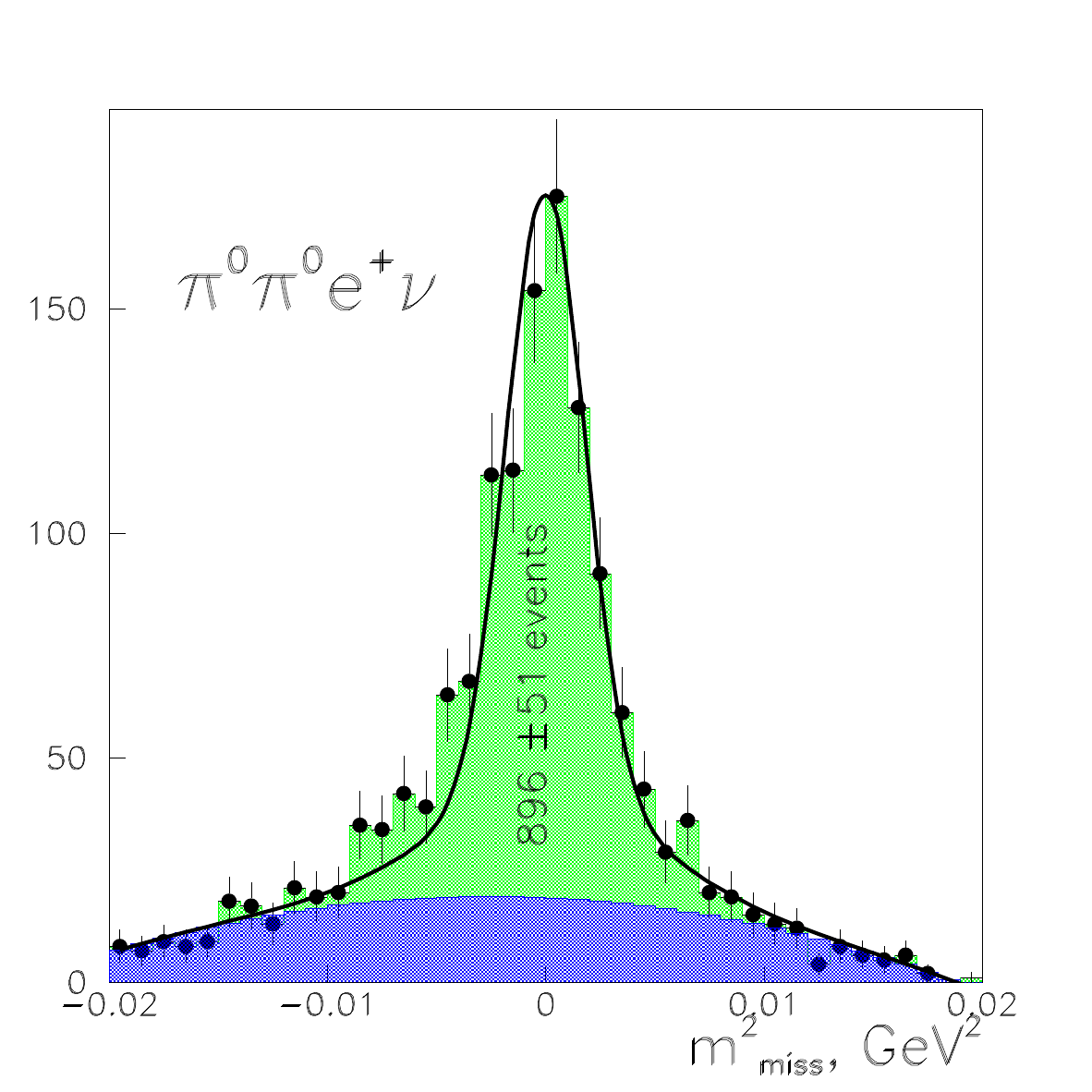}
\caption{\label{ke4} Observation of the $K^+ \to \pi^0\pi^0e^+\nu$ decay.
  Left: Energy balance in the event for the signal (hatched blue) and background (light);
  middle: fit of the signal MC missing mass squared spectrum to 2 Gaussian shapes;
  right: missing mass squared spectrum fit to the MC shape from fig.\ref{ke4}(middle) + $2^{nd}$ order polynomial.}
\end{center}
\end{figure*}

To suppress the major background source $K^+ \to \pi^+\pi^0\pi^0$ the cuts on the energy balance
$\Delta E < -2$~GeV (fig.\ref{ke4}(left))  and missing transverse momentum $P_T < 0.12$~GeV applied.
The cut on the missing energy in the $K^+$ rest frame  $E_{miss}^* $ also helps suppressing
backgrounds. The choice of $E_{miss}^*>0$ threshold is driven by the fact that in the decays with $\nu$
measured with good precision $E_{miss}^* \approx E_{\nu}^* > 0$.
Per MC, the negative values of $E_{miss}^* $ come mostly from picking the wrong combinations of $\gamma$s
misidentified as $\pi^0$s; finite resolution also contributes.
Clear peak in the missing mass is seen in fig.\ref{ke4}(right).
We fit this spectrum to the shape obtained from the fit of the signal MC spectrum in fig.\ref{ke4}(middle)
+ $2^{nd}$ order polynomial to account for the background. The integral of the MC-shape
 yields $896 \pm 51$ events of the decay.

\subsection{$K^+ \to \pi^0\pi^0\pi^0e^+\nu$}

\begin{figure}
  \centering
\includegraphics[width=0.5\linewidth]{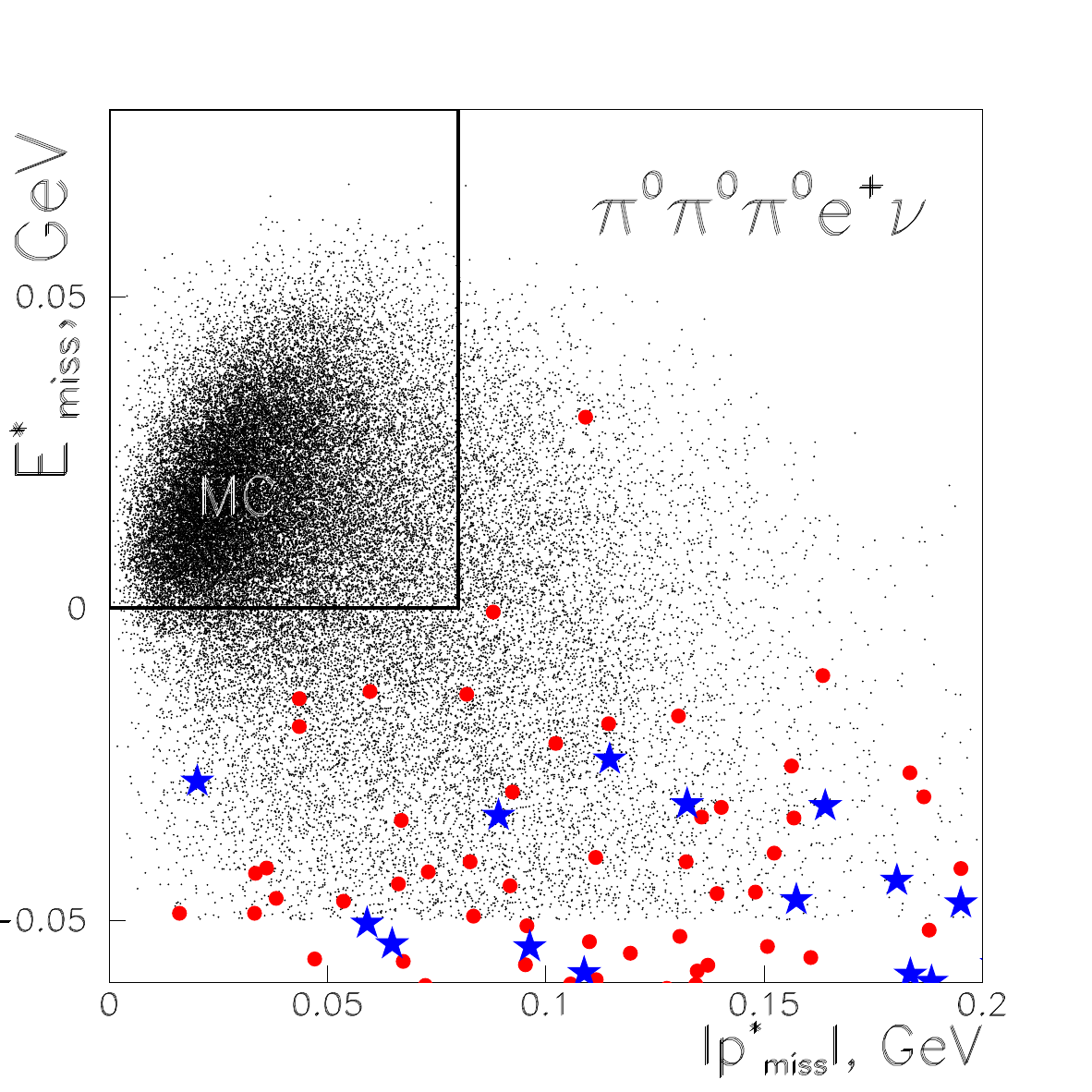}
\caption{\label{enuc3} Search for the $K^+ \to \pi^0\pi^0\pi^0e^+\nu$ decay:
  missing momentum modulus and missing energy in the $K^+$ rest frame, data (red bullets),
  MC backgound (blue stars) and MC signal (dots). No events pass selections shown with the rectangle.}
\end{figure}

We calculate missing energy $E_{miss}^*$ and missing momentum $p_{miss}^*$ in the $K^+$ rest frame;
they are not the same since we don't fix the missing mass.
Most of the candidates passing soft criteria fail the $E_{miss}^* >0$ test (fig.\ref{enuc3}).
Few remaining candidates are rejected by $|p_{miss}| < 0.08$~GeV requirement driven by kinematics of the decay:
the neutrino momentum can not exceed $p^*_{\nu} < \frac{M_K^2 - (3m_{\pi})^2}{2M_K} \approx 0.08$~GeV.
As one can see from fig.\ref{enuc3}, the candidates are not compatible with signal MC in any way.

\section{Branching ratios}

The detecting efficiency for every decay is obtained as the number of the events passing the
selection criteria divided by the total number of generated events of that decay.
The efficiency to the decay {\bf x},
$\epsilon(x)$, obtained this way, the number $n_x$ of the events detected in the expriment and
the table value $BR(K^+ \to \pi^0e^+\nu) = (5.07 \pm 0.04)\%$ are then useed to derive the BR:

\begin{equation}
  BR(x) = \frac{n_x}{n_{\pi^0e^+\nu}} \times \frac{\epsilon_{\pi^0e^+\nu}}{\epsilon_x} \times BR(\pi^0e^+\nu)
  \label{br-formula}
\end{equation}

$x=\pi^0\pi^0e^+\nu, \quad \pi^0\pi^0\pi^0e^+\nu, \quad n_{\pi^0\pi^0\pi^0e^+\nu} < 2.3$ for $90\%$CL.

The decays' matrix elements are calculated within the Standard Model framework:

\begin{equation}
  M \sim (\bar{e} \gamma_{\alpha} (1 + \gamma_5) \nu)H_{\alpha}
   \label{mtrx0}
\end{equation}

The Lorenz-invariance, Bose-statistics and negligible electron mass restrict the hadronic current $H_{\alpha}$ to the following
shapes:

\begin{eqnarray}
  H_{\alpha} &=& f_1 p_{\alpha} \quad \mbox{for} \quad  K^+ \to \pi^0e^+\nu,   \nonumber \\
  H_{\alpha} &=& f_1 (p_1 + p_2)_{\alpha} \quad  \mbox{for} \quad  K^+ \to \pi^0\pi^0e^+\nu,   \nonumber \\
  H_{\alpha} &=& f_1 (p_1 + p_2 + p_3)_{\alpha} + f_4 q_{\alpha}, \nonumber \\
    q &=& \frac{\{(p_1 \cdot p_2)p_3\}_{123}}{m_{\pi}^2} \quad  \mbox{for} \quad  K^+ \to \pi^0\pi^0\pi^0e^+\nu
   \label{mtrx1}
\end{eqnarray}

Here $p_i$ are the pion's momenta with \{\} meaning the symmetrization over 3 $\pi^0$s.
$f_{1,4}(m_{e\nu},m_{h})$ are the form-factors generally dependent on the masses of lepton and hadron
systems:

$ m_{e\nu}^2 = (k_e + k_{\nu})^2, m_h^2 = \left( \sum_i p_i \right)^2$.
For $f_1$ we use the phenomenological parameterization from \cite{Ke4-NA48}; it has little effect on detection efficiency.

\begin{figure} 
  \centering
\includegraphics[width=0.33\textwidth]{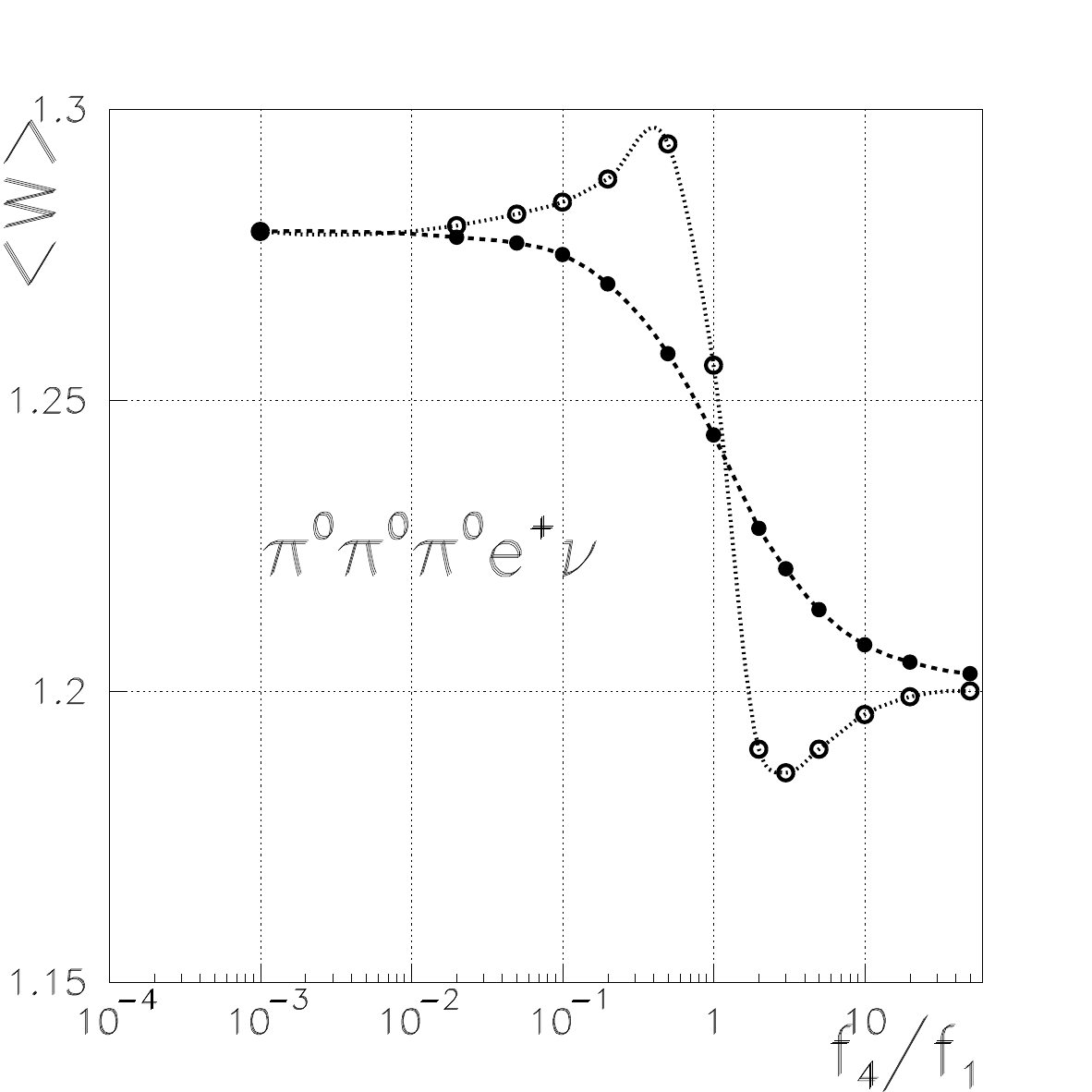}
\caption{\label{e3pi} Average weight of $K^+ \to \pi^0\pi^0\pi^0e^+\nu$ events passing selection criteria
  as function of $f_4/f_1$, open circles are for the case $f_4<0$.}
\end{figure}

There is no information on $f_4$. However, the MC simulation suggests that the effect of $f_4$ term
on the efficiency does not exceed $9\%$
(fig.\ref{e3pi}). The result quoted below is for the worst case scenario $f_4=-3f_1$. The branching ratios are listed in
Table \ref{br-tabl}.

\begin{table*}[!ht]
\caption{\label{br-tabl} Branching ratios.}
\begin{center}
\centering
\begin{tabular}{ |c|c|c|c|c|} 
 \hline
  Decay & Events observed & Efficiency & BR, this work & BR, PDG  \\ 
 \hline
  $K^+ \to \pi^0e^+\nu$ & $8.4 \times 10^6$ & $1.08 \times 10^{-2}$ & Used for normalization & $(5.07 \pm 0.04) \%$  \\ 
hard cuts &  &  &  &   \\ 
 \hline
 $K^+ \to \pi^0\pi^0e^+\nu$ & $896 \pm 51$ & $2.3 \times 10^{-3}$ & $(2.54 \pm 0.14) \times 10^{-5}$ &
 $(2.55 \pm 0.04) \times 10^{-5}$  \\ 
 \hline
  $K^+ \to \pi^0\pi^0\pi^0e^+\nu$ & 0 & $1.89 \times 10^{-3}$ & $<5.4 \times 10^{-8} ~ 90\%CL$ & $<3.5 \times 10^{-6} ~ 90\%CL$  \\ 
 \hline
\end{tabular}
\end{center}
\end{table*}

\section{Systematic errors}

The most sizeable uncertainty, up to $9\%$, comes from the unknown $f_4/f_1$ ratio. We quote the upper limit
under the most pessimistic assumption $f_4=-3f_1$, it will go down to

$BR(K^+ \to \pi^0\pi^0\pi^0e^+\nu)<5 \times 10^{-8} \quad 90\%CL$ if $f_4<<f_1$ (fig.\ref{e3pi}).
The background-free environment for $K^+ \to \pi^0\pi^0\pi^0e^+\nu$ decay observation allows for very soft
selection criteria rendering the efficiency $\epsilon_{\pi^0\pi^0\pi^0e^+\nu}$ calculation insensitive to possible
MC imperfections. \\
The agreement of the obtained $BR(K^+ \to \pi^0\pi^0e^+\nu)$ with the PDG value proves that the efficiencies
$\epsilon_{\pi^0e^+\nu},\epsilon_{\pi^0\pi^0e^+\nu}$ are correct to at least $51/896 \approx 6\%$ statistical error.

The effect of such an uncertainty on the upper limit is negligible.
If we want to set an upper limit for some value {\bf B} which is related to the observed number of events
{\bf n} through $n=\epsilon B$ and $\epsilon$ is known with $\pm \sigma_{\epsilon}$ error\cite{syst-eff} then
for the normal distribution of $\epsilon$ the probability to observe
n=0 events is given by the convolution:
  $  P_0 = \frac{1}{\sqrt{2\pi}\sigma_{\epsilon}}\int{\exp \left[ -(\epsilon+x)B -\frac{x^2}{2\sigma_{\epsilon}^2} \right] dx} $. 

The integration yields
  $   P_0 = e^{-A}, \quad   A = \epsilon B \left[1 - \left( \frac{\sigma_{\epsilon}}{\epsilon} \right)^2 \times \frac{\epsilon B}{2}
  \right] $.

  At $90\%$ CL $ \quad P_0  = 0.1,  \quad A \approx 2.3, $ \quad
  $  \epsilon B \approx 2.3 \left[ 1 +  1.15\left( \frac{\sigma_{\epsilon}}{\epsilon} \right)^2  \right]  \quad
    \left( \frac{\sigma_{\epsilon}}{\epsilon} <<1 \right). $

    Thus the correction to the upper limit is only {\bf quadratic} in $(\sigma_{\epsilon}/\epsilon)$,
    i.e. negligible for any reasonable estimate of $\epsilon$.
  
\section{Conclusion}

Three decays, $K^{+} \to \pi^{0}e^+\nu$, $K^{+} \to \pi^{0}\pi^{0}e^+\nu$ and $K^{+} \to \pi^{0}\pi^{0}\pi^{0}e^+\nu$
are studied by the OKA collaboration; the first one is used for the normalization. 
The observed $BR(K^{+} \to \pi^{0}\pi^{0}e^+\nu)=(2.54 \pm 0.14) \times 10^{-5}$ agrees with
the PDG value within statistical errors. No signal of $K^{+} \to \pi^{0}\pi^{0}\pi^{0}e^+\nu$ observed.
The upper limit set $BR(K^{+} \to \pi^{0}\pi^{0}\pi^{0}e^+\nu) < 5.4 \times 10^{-8} ~ 90\%CL$ is 65 times lower
than the one currently published by PDG. Due to the uncertainty in the matrix element the upper limit
is quoted under pessimistic assumption, it may be up to $9\%$ lower. The background-free environment for
the decay search is an important result of this work since
it allows to improve the upper limit linearly (rather than the square root) with statistics available.
This observation paves the road for the future high-statistics experiments.

{\small
\subsection*{Acknowledgements}

We express our gratitude to our colleagues in the accelerator department for the good performance of the U-70 during data taking; 
to colleagues from the beam department for the stable operation of the 21K beam line, including RF-deflectors, and to colleagues 
from the engineering physics department for the operation of the cryogenic system of the RF-deflectors.\\
This work was supported by the RSCF grant {\it{N\textsuperscript{\underline{\scriptsize o}}}22-12-0051}.

 -----------------------------------------------------------------

\end{document}